\magnification  \magstep1
\centerline {\bf  The Gluon Propagator in the Coulomb Gauge  }
\vskip 2cm
\centerline  { A. Andra\v si  }
\centerline  {\it 'Rudjer Bo\v skovi\' c' Institute, Zagreb, Croatia }
\vskip 4cm

\beginsection Abstract

We give the results for all the one-loop propagators, including finite parts, in 
the Coulomb gauge. In finite parts we find new non-rational functions in addition to the single logarithms of the Feynman gauge. Of course, the two gauges must agree for any gauge invariant function. We revise the manuscript hep-th/0311118v2 and Eur.Phys.J.C37, 307-313(2004) in accordance with the
notation and correct Feynman rules for the Coulomb gauge in Minkowski space
found in [16]. The high-energy behaviour of the proper two-point functions is
added in Appendix C.

\vskip  1cm
PACS: 11.15.Bt; 11.10.Gh
\vskip 0.5cm
Keywords: Coulomb gauge; Gluon propagator
\vskip 5cm
Electronic address: aandrasi@rudjer.irb.hr

\vfill \eject

\beginsection 1. Introduction

The non-covariant axial and Coulomb gauges have more direct physical interpretation than the covariant gauges because their propagators are
closely related to the polarization states of real spin-1 particles.
The relevant diagrams in the Coulomb gauge are not plagued by ghosts.
Also the time-time component of the gluon propagator provides a long-range
confining force [1], [2]. The Hamiltonian for non-Abelian gauge theory in
the Coulomb gauge has been known for some time in its continuum version [3].
The Coulomb gauge in the Hamiltonian formalism is manifestly unitary. The
main point in its favour is that problems concerned with the definition of the axial gauge integrals like
$$ \int d^4k{1\over{(n\cdot k)^2}}  \eqno(1)  $$
do not appear in the definition of integrals like
$$  \int d^4k {1\over{K^2}} ... \eqno(2)  $$
in the Coulomb gauge.
However, there are disadvantages. The naive Coulomb gauge Feynman rules in non-
Abelian gauge theory give rise to ambiguous integrals, in addition to the usual ultra-violet divergences [4]. At one loop order and above there are integrals like
$$ \int {{d^3P}\over{(2\pi)^3}}\int {{dp_0}\over{(2\pi)}} {{p_0}\over{p_0^2 - P^2 +i\eta}}\times {1\over{(P-K)^2}}. \eqno(3)  $$
There is no regularization procedure for the energy divergence in $ p_0 $ within the standard dimensional regularization 
scheme.
This integral and similar more complicated divergences in higher order diagrams have been the subject of the study [5], [6], where systematic cancellations have been found. However, no general proof exists that controls all divergences [7].
Formally such integrals are assigned value zero.
The Coulomb gauge has been extensively studied in the phase space formalism by 
D. Zwanziger [8] in the Euclidean space. The ultra-violet divergent parts of the proper two-point functions have been calculated and found to observe the Ward identities. In addition, a more powerful Ward identity holds in the Coulomb gauge than is available in covariant gauges. In this paper we give the results for the complete propagator to order $ g^2 $ including finite parts in Minkowski space.

\beginsection  2. The Coulomb gauge in the phase-space formalism

We use the phase-space formalism in order to avoid the ambiguous integrals like (3).
Let the generating functional of the Green's functions be
$$ Z(j,J) = \int d[f]d[A][J^{\mu}A_{\mu} + j^{\mu \nu}f_{\mu \nu}] \exp 
[-i\int d^4x L ], \eqno(4)  $$ where $ J $, $ j $ are sources, $ L $ the Lagrangian density and $ f $ and $ A $ are the fields [9].
$$ L = -{1\over 4}f^a_{\mu \nu}f^{a \mu \nu} + {1\over 2}f^{a \mu \nu}F^a_{\mu \nu}, \eqno(5) $$
and
$$ F^a_{\mu \nu} = \partial _{\mu}A^a_{\nu}-\partial _{\nu}A^a_{\mu} + gf^{abc}A^b_{\mu}A^c_{\nu}.
\eqno(6) $$
The Greek indices run from $ 0 $ to $ 3 $ and Latin indices denote spatial dimensions $ (i=1,2,3) $, $ a $, $ b $, $ c $,
are colour indices. Instead of setting the source $ j $ to zero as in Lagrangian formalism, we keep some components
of $ j $. We write out $ L $ as
$$ L = -{1\over 4}(f^a_{ij})^2 + {1\over 2}(f^a_{0i})^2 + {1\over 2}f^a_{ij}F^a_{ij} - f^a_{0i}F^a_{0i} \eqno(7) $$
and set $ j_{ij}=0 $ in order to perform the Feynman integral over $ f_{ij} $. We denote $ f_{0i} = E_i $.
The Lagrangian becomes
$$ L = -{1\over 4}(F^a_{ij})^2 - {1\over 2}(E^a_{i})^2 + E^a_{i}F^a_{0i} \eqno(8) $$
where
$$ E^a_{i}F^a_{0i} = E^a_{i}[\partial _0A^a_i - \partial _iA^a_0 + g f^{abc}A^b_0A^c_i]. \eqno(9) $$
The field $ E_i $ is the momentum conjugate to $ A_i $.
We have $ L $ expressed in terms of momenta and linear terms in derivatives. It is first order in time derivatives.
Adding the gauge fixing term
$$ {1\over{2\alpha}}(\partial _i A_i)^2 \eqno(10) $$
 we can deduce the propagators from the quadratic part of the Lagrangian. The propagator matrix is the inverse of the
matrix of these quadratic parts, and is given by the following $ 7\times 7 $ matrix, whose rows and columns are labeled
by $ A_1 $, $A_2 $, $A_3 $ ; $ A_0 $; $ E_1 $, 
$ E_2 $, $E_3 $ :
\vfill\eject

\hrule height 1pt
\vskip 2mm
\settabs\+ $A_iMMMM$ & $+{T_{ij}}/{k^2}-\alpha{L_{ij}}/{K^2}MMMM $ & +$1/{K^2}+\alpha {k_0^2}/{(K^2)^2}MMMM$ &+$ -{ik_0T_{in}}/{k^2}MMMM$ \cr
\+  & $ A_j $ & $A_0$ & $ E_n $   \cr 
\vskip 2mm
\hrule height 1pt \vskip 2mm
\+ $A_i$ & $ +{T_{ij}}/{k^2}-\alpha{L_{ij}}/{K^2} $ & $ {\alpha k_0K_i}/{(K^2)^2} $ & $ -{ik_0T_{in}}/{k^2}$ \cr
\vskip 2mm
\+ $A_0$ & $ \alpha {k_0K_j}/{(K^2)^2}$ & $1/{K^2} +\alpha{k_0^2}/{(K^2)^2}$ & $i{K_n}/{K^2} $ \cr
\vskip 2mm
\+ $ E_m$ & $+i{k_0T_{mj}}/{k^2}$ & $-i{K_m}/{K^2}$ & $+{T_{mn}K^2}/{k^2}$ \cr
\vskip 2mm
\hrule height 1pt
\vskip 5mm
where
$$ T_{ij}\equiv \delta_{ij}-L_{ij},{\hskip 1cm }  L_{ij}\equiv {K_iK_j}/{K^2}, $$
$$ k^2 = k_0^2-K^2. \eqno(11) $$
The Coulomb gauge propagators are obtained by setting $ \alpha = 0 $.

\beginsection  3. The proper two-point functions

The method of evaluation of Coulomb gauge integrals is explained in Appendix A.
Here we list the results. The constants used throughout the paper are
$$ \epsilon = 4-d  \eqno(12) $$
where $ d $ is the dimension of space-time and the coupling parameter is
$$ c\delta_{ab}= {{ig^2}\over{16\pi^2}}C_G\delta_{ab} \eqno(13) $$.

\beginsection  The transverse gluon two-point function

There are two non-vanishing graphs contributing to the transverse gluon propagator.
The graph shown in Fig.1 gives
$$\Gamma_1^{A_iA_j}= c\Gamma({{\epsilon}\over 2}){({{K^2}\over{\mu^2}})}^{-{{\epsilon}\over 2}}\times 2^{-\epsilon}
(1+{{77}\over{30}}\epsilon) $$  $$ \times \{{{12}\over{15}}K^2\delta_{ij} -{{16}\over{15}}K_iK_j +
{{4}\over{15}}\epsilon K_iK_j \} \eqno(14) $$
The graph in Fig.2 contributes
$$ \Gamma_2^{A_iA_j} = c(MK_iK_j + NK^2\delta_{ij})  \eqno(15) $$
where

\vfill \eject

$$ M={{31}\over{15}}\Gamma({{\epsilon}\over 2})-{8\over 3}\ln({{-k^2-i\eta}\over{\mu^2}})
+{3\over{5}}\ln{{K^2}\over{\mu^2}} $$ $$ +{1\over 4}[-{{K^4}\over{k_0^2}}+18{{k_0^2k^2}\over{K^2}}
+9{{k_0^2k^4}\over{K^4}}+k_0^2]\times D $$ $$ +{1\over 4}[-{{K^3}\over{k_0^3}}+18{{k_0k^2}\over{K^3}}
+9{{k_0k^4}\over{K^5}}+{{k_0}\over{K}}]\ln{{k_0+K-i\eta}\over{k_0-K+i\eta}}\times \ln{{K^2}\over{(-k^2-i\eta)}} $$
$$ -{1\over{2K^2}}[3k_0^2-5K^2+9{{k_0^4}\over{K^2}}-{{K^4}\over{k_0^2}}]\ln{{K^2}\over{(-k^2-i\eta)}} $$
$$ -{{(\ln 2-1)}\over{K^2}}[15k^2+6k_0^2+9{{k^4}\over{K^2}}-{3\over 4}{{K^4}\over{k_0^2}}-{{K^4}\over{4k^2}}+{7\over{15}}K^2] $$
$$ +{{k_0^2}\over{K^2}}+{{K^2}\over{k_0^2}}+{{K^2}\over{4k^2}}+6+{{11}\over{18}}-{8\over{15}}-{1\over{25}} \eqno(16) $$

$$ N=-{1\over{3K^2}}(k_0^2+{{27}\over 5}K^2)\Gamma({{\epsilon}\over 2})+{1\over{3K^2}}(k_0^2+8K^2)
\ln({{-k^2-i\eta}\over{\mu^2}})-{{13}\over{15}}\ln{{K^2}\over{\mu^2}} $$
$$+{1\over 4}[{{K^4}\over{k_0^2}}-{{k_0^2k^2}\over{K^2}}(14+{{K^2}\over{k^2}}+3{{k^2}\over{K^2}})]\times D $$
$$+{1\over 4}[{{K^3}\over{k_0^3}}-{{k_0k^2}\over{K^3}}(14+{{K^2}\over{k^2}}+3{{k^2}\over{K^2}})]
\ln{{k_0+K-i\eta}\over{k_0-K+i\eta}}\times \ln{{K^2}\over{(-k^2-i\eta)}} $$ 
$$+ {1\over{2K^2}}(9k^2+6k_0^2+K^2+3{{k^4}\over{K^2}}-{{K^4}\over{k_0^2}})\ln{{K^2}\over{(-k^2-i\eta)}} $$
$$+{{(\ln 2-1)}\over{K^2}}(9k^2+6k_0^2+3{{k^4}\over{K^2}}+K^2-{3\over 4}{{K^4}\over{k_0^2}}-{{K^4}\over{4k^2}}-{{26}\over{15}}K^2) $$
$$-{{K^2}\over{k_0^2}}-{{K^2}\over{4k^2}}-{8\over 9}{{k_0^2}\over{K^2}}+{2\over 9}+{{27}\over{25}} \eqno(17) $$

The non-rational structure $ D $ which appears in the results for the proper two-point functions is in the integral form

$$ D = \int_{0}^{1} dx {{x^{-{1\over 2}}}\over {k_0^2-x(K^2-i\eta)}}\ln(1-x). \eqno(18) $$

\vfill \eject

\beginsection  In the region $ k_0>K $

$$ D={1\over{k_0K}}\{Li_2({{k_0-K+i\eta}\over{k_0+K-i\eta}}) -Li_2({{k_0+K-i\eta}\over{k_0-K+i\eta}})  $$
$$ +\ln{{k_0+K-i\eta}\over{k_0-K+i\eta}}\times \ln{{k^2+i\eta}\over{K^2}}-i\pi \ln{{k_0+K-i\eta}\over{k_0-K+i\eta}}\}.
\eqno(19) $$

\beginsection  In the region $ K>k_0 $

$$ D={1\over{k_0K}}\{Li_2({{K-k_0-i\eta}\over{K+k_0-i\eta}})-Li_2({{K+k_0-i\eta}\over{K-k_0-i\eta}})  $$
$$+\ln{{K+k_0-i\eta}\over{K-k_0-i\eta}}\times \ln({{-k^2-i\eta}\over{k_0^2}})+i\pi \ln{{K+k_0-i\eta}\over{K-k_0-i\eta}}\}
$$ $$ -{2\over{k_0K}}[Li_2(-{{k_0}\over{K-i\eta}})-Li_2({{k_0}\over{K-i\eta}})] +{{i\pi}\over{k_0K}}
\ln{{K^2}\over{(-k^2-i\eta)}}  \eqno(20) $$
where
$$ Li_2(x)=-\int_{0}^{x}{{\ln (1-z)}\over {z}} dz \eqno(21)  $$
is the Spence function and $ k_0 $ and $ K $ are the lengths of the respective vectors. 
The two expressions for $ D $ in (19) and (20) are connected as analytic continuations
of each other with the relation (B1).

\beginsection  The $ A_iA_0 $ transition

The whole contribution to the $ A_iA_0 $ transition to order $ g^2 $ comes from the graph in Fig.3a.
$$ \Gamma ^{A_iA_0} = ck_0K_i\times Z \eqno(22) $$
$$ Z = -{1\over 3}\Gamma({{\epsilon}\over 2})({{-k^2-i\eta}\over{\mu^2}})^{-{{\epsilon}\over 2}} $$
$$+{{k^2}\over{2K^2}}(2k_0^2+k^2)\times D $$
$$+{{k^2}\over{2k_0K^3}}(2k_0^2+k^2)\ln{{k_0+K-i\eta}\over{k_0-K+i\eta}}\times \ln{{K^2}\over{(-k^2-i\eta)}} $$
$$ -3{{k^2}\over{K^2}}\ln{{K^2}\over{(-k^2-i\eta)}}-6{{k^2}\over{K^2}}\ln 2 +6{{k_0^2}\over{K^2}}-{{53}\over 9} 
\eqno(23) $$
The graph shown in Fig.3b contains integrals like the one in (3). Formally such integrals are assigned value zero.

\vfill \eject

\beginsection  The time-time component of the two-point function

Two graphs contribute to the $ A_0A_0 $ function. The graph shown in Fig.4a gives
$$ \Gamma_a^{A_0A_0} = c\{\Gamma({{\epsilon}\over 2})({{-k^2-i\eta}\over{\mu^2}})^{-{{\epsilon}\over 2}}\times
({1\over 2}k_0^2+{5\over 6}K^2+{{\epsilon}\over{12}}k^2+{{\epsilon}\over 6}k_0^2+{{17}\over{18}}\epsilon K^2) $$
$$ -2^{-\epsilon}({5\over 3}+{{28}\over 9}\epsilon)\Gamma({{\epsilon}\over 2})K^2({{K^2}\over{\mu^2}})^{-{{\epsilon}\over2}}
$$ $$ +{1\over 2}k^4\times D $$ 
$$+{{k^4}\over{2k_0K}}\ln{{k_0+K-i\eta}\over{k_0-K+i\eta}}\times \ln{{K^2}\over{(-k^2-i\eta)}} $$
$$-k_0^2\ln{{K^2}\over{(-k^2-i\eta)}}-2(\ln 2 -1)k_0^2 \} \eqno(24) $$
The graph in Fig.4b contributes
$$ \Gamma_b^{A_0A_0}=c\{\Gamma({{\epsilon}\over 2})({{-k^2-i\eta}\over{\mu^2}})^{-{{\epsilon}\over 2}}\times
({1\over 3}K^2-{1\over 2}k^2-{{\epsilon}\over 4}k^2+{{11}\over{18}}\epsilon K^2) $$
$$ -{1\over 3}\Gamma({{\epsilon}\over 2})K^2({{K^2}\over{\mu^2}})^{-{{\epsilon}\over 2}}-K^2({{10}\over 9}
-{{2\ln 2}\over 3}) $$ $$ +k^2k_0^2\times D $$ 
$$+{{k_0k^2}\over K}\ln {{k_0+K-i\eta}\over{k_0-K+i\eta}}\times \ln{{K^2}\over{(-k^2-i\eta)}} $$
$$-(2k^2+K^2)\ln{{K^2}\over{(-k^2-i\eta)}}-2(2k^2+K^2)(\ln 2 -1)\} \eqno(25) $$
We can verify that the complete proper two-point functions satisfy the 't Hooft identity [10]
$$ k_0^2\Gamma^{A_0A_0} -2k_0K_i\Gamma^{A_iA_0} + K_iK_j\Gamma^{A_iA_j} = 0 \eqno(26) $$
and the stronger Zwanziger identity [8]
$$ k_0\Gamma^{A_0A_0}=K_i\Gamma^{A_iA_0}. \eqno(27) $$
The remaining graphs contain the conjugate field $ E_i $ as the external leg.

\beginsection $ E_iA_j $ graph

The graph in Fig.5 vanishes as the energy divergence.

\beginsection $ E_iA_0 $ graph

The graph in Fig.6 gives
$$\Gamma^{E_iA_0} = -c({{K^2}\over{\mu^2}})^{-{{\epsilon}\over 2}}\{{4\over 3}-2^{-\epsilon}
\Gamma({{\epsilon}\over 2})({2\over 3}+{{13\epsilon}\over 9})\}\times(2iK_i) \eqno(28) $$

\beginsection   $ E_iE_j $ graph

The graph in Fig.7 contributes
$$ \Gamma^{E_iE_j}= 2c({{K^2}\over{\mu^2}})^{-{{\epsilon}\over 2}}\times \{2^{-\epsilon}({2\over 3}+
{{13\epsilon}\over 9})\Gamma({{\epsilon}\over 2})\delta_{ij}-{4\over 3}{{K_iK_j}\over{K^2}}\} \eqno(29) $$

\beginsection  4. The gluon propagator to order $ g^2 $

We form the $ 7\times 7 $ matrix of  free and order $ g^2 $ proper two-point functions. The inverse of this
matrix gives the gluon propagator to order $ g^2 $.

\beginsection  The $ A_0A_0 $ propagator

The time-time component of the gluon propagator to order $ g^2 $ is [11]
$$ D^{A_0A_0}={1\over{K^4}}[\Gamma^{A_0A_0}-iK_n\Gamma^{A_0E_n}]+{{iK_m}\over{K^4}}[\Gamma^{E_mA_0}
-iK_n\Gamma^{E_mE_n}]  \eqno(30) $$
or explicitly
$$ D^{A_0A_0} =c(K^2)^{-2}\times \{{{11}\over 3}\Gamma({{\epsilon}\over 2})K^2
-{5\over 3}K^2\ln{{(-k^2-i\eta)}\over{\mu^2}}-2K^2\ln{{K^2}\over{\mu^2}} $$
$$+{1\over 2}k^2(k^2+2k_0^2)\times D $$
$$+{{k^2}\over{2k_0K}}(k^2+2k_0^2)\ln {{k_0+K-i\eta}\over{k_0-K+i\eta}}\times \ln{{K^2}\over{(-k^2-i\eta)}} $$
$$-(3k_0^2-K^2)\ln{{K^2}\over{(-k^2-i\eta)}}-(6k_0^2+2K^2)(\ln 2-1)+{{13}\over 9}K^2 \} \eqno(31) $$
The ultraviolet divergent part of (31) gives the gauge invariant Coulomb field renormalization factor [12].

\beginsection   The $A_iA_j $ propagator

The transverse gluon propagator to order $ g^2 $ is
$$ D^{A_iA_j}={1\over{k^4}}T_{im} \Gamma^{A_mA_n} T_{nj}+{{k_0^2}\over{k^4}}T_{im} \Gamma^{E_mE_n}T_{nj}  \eqno(32) $$

\vfill  \eject

or explicitly
$$ D^{A_iA_j} ={c\over{k^2+i\eta}}(\delta_{ij}-{{K_iK_j}\over{K^2}})\times \{\Gamma({{\epsilon}\over 2})
-{4\over 3}\ln{{K^2}\over{\mu^2}}+{1\over 3}\ln ({{-k^2-i\eta}\over{\mu^2}}) $$
$$+3{{K^2}\over{k^2}}\ln({{-k^2-i\eta}\over{\mu^2}}) $$
$$-{{K^2}\over 4}[{{K^2+k_0^2}\over{k_0^2}}+{{k_0^2}\over{K^2}}(14+3{{k^2}\over{K^2}})]\times D $$
$$-{K\over {4k_0}}[{{K^2+k_0^2}\over{k_0^2}}+{{k_0^2}\over{K^2}}(14+3{{k^2}\over{K^2}})]
\ln{{k_0+K-i\eta}\over{k_0-K+i\eta}}\times \ln{{K^2}\over{(-k^2-i\eta)}} $$
$$+{1\over 2}(15+3{{k^2}\over{K^2}}+{{K^2}\over{k_0^2}})\ln{{K^2}\over{(-k^2-i\eta)}}$$
$$+{1\over{k^2}}(\ln 2-1)(9k^2+{{10}\over 3}k_0^2+3{{k^4}\over{K^2}}
-{{3K^4}\over{4k_0^2}}-{{K^4}\over{4k^2}}-{7\over 3}K^2 ) $$
$$+{{20}\over 9}{{k_0^2}\over{k^2}}-{{K^4}\over{k_0^2k^2}}
-{{K^4}\over{4k^4}}+{{K^2}\over{k^2}}(3+{2\over 9}+{{34}\over{75}}) \eqno(33) $$

\beginsection  5. The Slavnov-Taylor identity

Although ghosts are absent from the S-matrix elements they are necessary to formulate the Slavnov-Taylor identities
[13],[14]. Diagramatically they are shown for the self-energy in Fig.8.
Algebraically they are
$$ k_0\Gamma^{A_0A_j}-K_i\Gamma^{A_iA_j}=(K^2\delta_{ij}-K_iK_j)\Gamma^{CA_i}  \eqno(34) $$
The diagrams involving ghost-source vertices on the right-hand side are shown in Fig.9a and Fig.9b.
The diagram in Fig.9a vanishes as the energy divergence in $ p_0 $.
The diagram in Fig.9b contributes
$$ \Gamma^{CA_i} =-2c({{K^2}\over{\mu^2}})^{-{{\epsilon}\over 2}}K_i\{-{4\over 3}+2^{-\epsilon}
\Gamma({{\epsilon}\over 2})({2\over 3}+{{13\epsilon}\over 9})\}  \eqno(35) $$
so the identity is satisfied trivially as implied by (26) and (27).

\beginsection 6. Discussion
       
We have checked the consistency of the Coulomb gauge to order $ g^2 $ including finite parts.
The time-time component of the gluon propagator in the Coulomb gauge is believed to provide a long-range confining
force. There are two interesting limits of eq.(31). In the Zwanziger picture [8] $ g^2D_{00} $ gives the instantaneous
part $ V_Z(R) $, which is called the color-Coulomb potential. (Here $ D_{00} $ is the time-time component of the
gluon propagator). The instantaneous color-Coulomb potential $ V_Z(R) $ at large $ R $ may serve as an order parameter.
$$ K_{Coul} \equiv \lim_{R \rightarrow \infty}{{V_Z(R)}\over R}  \eqno(36) $$
A non-zero value of $ K_{Coul} $ would be the signal for color confinement. The potential is separated out in momentum
space by
$$ V_Z(K) = \lim_{k_0 \rightarrow \infty}g^2D^{A_0A_0}(k_0,K)  \eqno(37) $$
where we have written $ V_Z(K) $ for the Fourier transform of $ V_Z(R) $.
The limit $ k_0\rightarrow \infty $ of eq.(31) is
$$ \lim_{k_0 \rightarrow \infty} D^{A_0A_0}(k_0,K) ={c\over{K^2}}\{{{11}\over 3}\Gamma({{\epsilon}\over 2})
-{{11}\over 3} \ln{{K^2}\over{\mu^2}} -{1\over 3}i\pi -8\ln 2+{{79}\over 9}-{2\over 3}\ln{K\over{k_0}}\}\eqno(38) $$
and it is not independent of $ k_0 $. The dominant term agrees with (38) and equation (37)
of A. Cucchieri and D. Zwanziger [15]. Their $ I_2 $ indeed vanishes in the limit $ k_0 \rightarrow \infty $.

Although the limit as $ k_0 \rightarrow \infty $ is not finite, A. Cucchieri and D. Zwanziger [15] have argued
that an unambiguous instantaneous part may be defined by using renormalization group arguments.

The limit $ k_0\rightarrow 0 $ is naturally related to the definition of the quark-antiquark potential. It follows from
considering a rectangular Wilson loop with sides of length $ T $ in the time direction (where $ T\rightarrow \infty $)
and $ L $ in the space direction. In the Coulomb gauge the main contribution comes from the $ D_{00} $ component
of the propagator (where $ k_0\rightarrow 0 $) attached to the two time-like sides. The $ k_0\rightarrow 0 $ limit of eq.(31) is
$$ \lim_{k_0\rightarrow 0}D^{A_0A_0}(k_0, K)={c\over{K^2}}\{{{11}\over 3}\Gamma({{\epsilon}\over 2})
-{{11}\over 3}\ln{{K^2}\over{\mu^2}}+{{31}\over 9}\} \eqno(39) $$
leading to the quark-antiquark potential
$$ V(R)=-2\pi^2g_r^2(\mu){1\over R}\{1+{{g^2C_G}\over{16\pi^2}}[{{31}\over 9}+{{11}\over 3}\gamma +{{11}\over 3}\ln(\mu R)^2]\} \eqno(40)$$
where $ \gamma $ is the Euler's constant, $ g_r(\mu) $ is the running coupling constant.
 If we assume the relation
$$ R\times \mu = 1 \eqno(41) $$
$g_r(\mu) $ becomes $ R $ dependent. We suppose that the exact $ g_r({1\over R}) $ tends to zero as
 $ R\rightarrow 0 $
and $ g_r({1\over R}) \rightarrow \infty $ for $ R\rightarrow \infty $.

\beginsection  Acknowledgements

It is a privilege to thank Prof. J. C. Taylor who gave me invaluable advice and encouragement which made this work
possible. This work was supported by the Ministry of Science and Technology of the Republic of Croatia under
Contract No. 0098003.

\beginsection Appendix A

We use two basic integrals for evaluation of the Coulomb gauge integrals.
$$ A=\int d^{4-\epsilon}p{1\over{p^2+i\eta}}\cdot {1\over{(k-p)^2+i\eta}}$$ $$ ={1\over 2}i\pi\int_{0}^{1}dy
\int d^{3-\epsilon}P\{P^2-2P\cdot Ky-yk^2+y^2k_0^2-i\eta\}^{-{3\over 2}} \eqno(A1) $$
$$ B=\int d^{4-\epsilon}p{{p_0}\over{(p^2+i\eta)[(k-p)^2+i\eta]}} $$
$$={1\over 2}i\pi k_0\int_{0}^{1}y dy\int d^{3-\epsilon}P\{P^2-2P\cdot K y-yk^2+y^2k_0^2-i\eta \}^{-{3\over 2}}
\eqno(A2) $$ 
As an example we evaluate the integral
$$ X_{ij}=\int d^{4-\epsilon}p{{p_0}\over{p^2+i\eta}}\cdot {1\over{(k-p)^2+i\eta}}\cdot{{P_iP_j}\over{P^2}}\eqno(A3)$$
Applying (A2)
$$X_{ij}={1\over 2}i\pi k_0\int_{0}^{1} ydy\int d^{3-\epsilon}P{{P_iP_j}\over{P^2}}\cdot 
{1\over{[P^2-2P\cdot Ky-yk^2+y^2k_0^2-i\eta]^{{3\over 2}}}} \eqno(A4) $$
Combining the denominators with the Feynman parameter $ x $
$$ X_{ij}=ik_0\pi^{1\over 2}\Gamma({5\over 2})\int_{0}^{1}dx x^{1\over 2}\int_{0}^{1} ydy \int d^{3-\epsilon}P
{{P_iP_j}\over{[P^2-2P\cdot Kxy-xyk^2+y^2xk_0^2-i\eta x]^{5\over 2}}} \eqno(A5) $$
Now it is easy to perform the $ d^{3-\epsilon}P $ and integration over the parameter $ y $ giving $ X_{ij} $
in the integral form.
$$ C^{-1}X_{ij}={1\over 6}\Gamma({{\epsilon}\over 2})(K^2)^{-{{\epsilon}\over 2}}\delta_{ij}-{1\over 3}
{{K_iK_j}\over{K^2}}+{1\over 3}\delta_{ij}({{13}\over 6}-\ln 2) $$ 
$$+({1\over 4}\delta_{ij}-{{K_iK_j}\over{K^2}})k^2\{\int_{0}^{1}dx{{x^{1\over 2}}\over{k_0^2-x(K^2-i\eta)}}
+k^2\int_{0}^{1}dx{{x^{1\over 2}}\over{[k_0^2-x(K^2-i\eta)]^2}}\ln{{K^2(1-x)}\over{(-k^2-i\eta)}}\} $$
$$+{{K_iK_j}\over{K^2}}k_0^2\{{1\over 2}\int_{0}^{1}dx{{x^{1\over 2}}\over{k_0^2-x(K^2-i\eta)}}
+k^2\int_{0}^{1} dx{{x^{1\over 2}}\over{[k_0^2-x(K^2-i\eta)]^2}} $$
$$+k^4\int_{0}^{1}dx{{x^{1\over 2}}\over{[k_0^2-x(K^2-i\eta)]^3}}\ln {{(1-x)K^2}\over{(-k^2-i\eta)}}\} $$
where 
$$ C=ik_0\pi^{{4-\epsilon}\over2}. \eqno(A6) $$ The integrals in (A6) are
$$\int_{0}^{1}dx{{x^{1\over 2}}\over{k_0^2-x(K^2-i\eta)}}=-{2\over{K^2}}+{{k_0}\over{K^3}}
\ln{{k_0+K-i\eta}\over{k_0-K+i\eta}} \eqno(A7) $$
$$ \int_{0}^{1}dx{{x^{1\over 2}}\over{[k_0^2-x(K^2-i\eta)]^2}}={1\over{k^2K^2}}-{1\over{2k_0K^3}}
\ln{{k_0+K-i\eta}\over{k_0-K+i\eta}} \eqno(A8) $$
$$ \int_{0}^{1}dx{{x^{1\over 2}}\over{[k_0^2-x(K^2-i\eta)]^3}}={1\over{2K^2k^4}}-{1\over{4k_0^2K^2k^2}}
-{1\over{8k_0^3K^3}}\ln{{k_0+K-i\eta}\over{k_0-K+i\eta}} \eqno(A9)  $$
$$ T=\int_{0}^{1}dx{{x^{1\over 2}}\over{k_0^2-x(K^2-i\eta)}}+k^2\int_{0}^{1}dx 
{{x^{1\over 2}}\over{[k_0^2-x(K^2-i\eta)]^2}}\ln{{K^2(1-x)}\over{(-k^2-i\eta)}} $$ $$ ={2\over{K^2}}(\ln 2-1)
+[{1\over{K^2}}-{{k^2}\over{2k_0K^3}}\ln{{k_0+K-i\eta}\over{k_0-K+i\eta}}]\times \ln{{K^2}\over{(-k^2-i\eta)}}
-{{k^2}\over{2K^2}}D  \eqno(A10) $$
$$ E=\int_{0}^{1}dx{{x^{1\over 2}}\over{[k_0^2-x(K^2-i\eta)]^3}}\ln (1-x)$$ $$={{k_0^2+K^2}\over{2k_0^2K^2k^4}}\ln 2
-{1\over{2K^2k^4}}-{1\over{2Kk_0k^4}}\ln{{k_0+K-i\eta}\over{k_0-K+i\eta}}-{1\over{8K^2k_0^2}}D \eqno(A11) $$
where $ D $ was defined in (18) and the explicit result given in (19) and (20).

\beginsection Appendix B

The expressions for $ D $ in (19) for $ k_0>K $ and in (20) for $ K>k_0 $ ought to be connected by analytic 
continuation. It is easy to see this happens using the following relation between the Spence functions.
$$ Li_2({{x-1+i\eta}\over{x+1-i\eta}})-Li_2({{x+1-i\eta}\over{x-1+i\eta}})+Li_2({{1+x-i\eta}\over{1-x-i\eta}})
-Li_2({{1-x-i\eta}\over{1+x-i\eta}}) $$ $$ +2Li_2(-x-i\eta)-2Li_2(x+i\eta)+\ln{{x+1-i\eta}\over{x-1+i\eta}}
\times \ln(x^2) $$ $$-i\pi \ln{{x+1-i\eta}\over{x-1+i\eta}}+i\pi \ln(x^2) +\pi^2=0  \eqno(B1) $$
where
$$ Li_2(x)=-\int_{0}^{x}{{\ln(1-z)}\over z}dz  \eqno(B2) $$

\beginsection Appendix C

We list the high-energy limits of some functions.

$$ \lim_{k_0 \rightarrow \infty} D={4\over{k_0^2}} (\ln 2 -1) +
{{K^2}\over{k_0^4}}({4\over 3} \ln 2 -{{16}\over 9}) \eqno(C1) $$

The expressions for $ M $ in (16) and $ N $ in (17) have the following
high-energy limits.

$$ \lim_{k_0 \rightarrow \infty} M={{31}\over{15}}\Gamma({{\epsilon}\over 2})
-{8\over 3}\ln ({{-k^2-i\eta}\over{ \mu^2}})+{3\over 5}\ln ( {{K^2}\over{\mu^2}})
-{3\over 2} \ln {{K^2}\over {(-k^2-i\eta)}} \eqno(C2) $$

$$ \lim_{k_0 \rightarrow \infty} N=-{{k_0^2}\over{3K^2}}\Gamma({{-k^2-i\eta}\over{\mu^2}})^{-{{\epsilon}\over 2}}
-{5\over 9}{{k_0^2}\over{K^2}} \eqno(C3) $$

$ Z $ appearing in (23) has the high-energy limit

$$ \lim_{k_0 \rightarrow \infty}Z=-{1\over 3}\Gamma({{\epsilon}\over 2})
({{-k^2-i\eta}\over{\mu^2}})^{-{{\epsilon}\over 2}}-{5\over 9} \eqno(C4) $$

$$ \lim_{k_0 \rightarrow \infty}\Gamma^{A_0A_0}=
-{1\over 3}K^2\Gamma({{\epsilon}\over 2})({{-k^2-i\eta}\over{\mu^2}})^{-{{\epsilon}\over 2}}
 -{5\over 9}K^2  \eqno(C5) $$

$$ \lim_{k_0 \rightarrow \infty}D^{A_iA_j}={c\over{k^2+i\eta}}
(\delta_{ij}-{{K_iK_j}\over{K^2}})\{-{1\over 3}\Gamma({{\epsilon}\over 2})
({{-k^2-i\eta}\over{\mu^2}})^{-{{\epsilon}\over 2}}
+{4\over 3}\Gamma({{\epsilon}\over 2})({{K^2}\over{\mu^2}})^
{-{{\epsilon}\over 2}}-{8\over 3}(\ln 2-1)+{{23}\over 9}\} \eqno(C6) $$

\beginsection  References

[1] N. Gribov, Nucl. Phys. {\bf B139} (1978) 1;

\noindent
[2] D. Zwanziger, Nucl. Phys. {\bf B485} (1997) 185;

\noindent
[3] N. Christ and T. D. Lee, Phys. Rev. {\bf D22} (1980) 939;

\noindent
[4] H. Cheng and E. C. Tsai, Phys. Lett. {\bf B176} (1986) 130;

\noindent
[5] P. Doust and J. C. Taylor, Phys. Lett. {\bf197} (1987) 232;

\noindent
[6] P. Doust, Ann. of Phys. {\bf 177} (1987) 169;

\noindent
[7] J. C. Taylor, in Physical and Nonstandard Gauges, Proceedings,

\line{\hskip 0.5cm Vienna, Austria 1989, P.Gaigg, W. Kummer and M. Schweda (Eds.); \hfill}

\noindent
[8] D. Zwanziger, Nucl. Phys. {\bf B518 } (1998) 237;

\noindent
[9] J. C. Taylor, private communication;

\noindent
[10] G. 't Hooft, Nucl. Phys. {\bf B35} (1971) 167;

\noindent
[11] A. Andra\v si, Europhys. Lett. {\bf Vol.66}, (3), (2004) 338;

\noindent
[12] J. Frenkel and J. C. Taylor, Nucl. Phys. {\bf B109} (1976) 439;

\noindent
[13] J. C. Taylor, Nucl. Phys. {\bf B33} (1971) 436;

\noindent
[14] A. Slavnov, Theor. and Math. Phys. {\bf 10} (1972) 99;

\noindent
[15] A. Cucchieri and D. Zwanziger, Phys. Rev. {\bf D65} (2002) 014002 

\noindent
[16] A. Andra\v si and J. C. Taylor, Ann. of Phys. {\bf 324} (2009) 2179

\vskip 1cm

\beginsection Figure Captions

Fig.1. The transverse gluon self-energy graph. The dashed line is the transverse gluon $ A_i $, the dotted line
represents the instantaneous Coulomb field $ A_0 $ and the continuous line is the $ E_i $ field conjugate to the
transverse field $ A_i $.
\vskip 0.5cm
\noindent
Fig.2. The transverse gluon self-energy graph. The dashed line is the transverse gluon field $ A_i $.
\vskip 0.5cm
\noindent
Fig.3a. The $ A_iA_0 $ two-point function. The dotted line is the instantaneous Coulomb field $ A_0 $, the dashed 
line represents the transverse field $ A_i $ and the solid line is the conjugate field $ E_i $.
\vskip 0.5cm
\noindent
Fig.3b. The $ A_iA_0 $ two-point function. The graph is suppressed as an energy divergence.
\vskip 0.5cm
\noindent
Fig.4a. The time-time component of the gluon self-energy to order $ g^2 $. The dotted lines represent the
instantaneous Coulomb field $ A_0 $. The continuous line is the $ E_i $ field conjugate to the transverse field
$ A_i $. The propagators inside the loop are the $ E_iA_j $ transitions specific to the Coulomb gauge.
\vskip 0.5cm
\noindent
Fig.4b. Self-energy graph to order $ g^2 $. The dotted line is the $ A_0 $ field, the dashed line is the transverse 
propagator and the solid line is the $ E_iE_j $ propagator.
\vskip 0.5cm
\noindent
Fig.5. The transition between the transverse gluon field and its conjugate field $ E_i $.
\vskip 0.5cm
\noindent
Fig.6. The transition between the Coulomb field $ A_0 $ and the conjugate field $ E_i $.
\vskip 0.5cm
\noindent
Fig.7. The conjugate field self-energy.
\vskip 0.5cm
\noindent
Fig.8. The Slavnov-Taylor identity for self-energy graphs. The wavy lines stand for Yang-Mills particles and double
lines for ghosts. The symbol on the left wavy line stands for the replacement of a polarization vector
$ e_{\mu}(k) $ by $ k_{\mu} $ and $ k^2 $ need not be zero. The cross denotes the action of the tensor
$ (k_{\mu}k_{\nu}-k^2\delta_{\mu \nu}) $. The circle represents the set of all relevant Feynman graphs.
\vskip 0.5cm
\noindent
Fig.9a. Diagram with an open ghost line. The source $ v_n $ of the $ E_m $ field has the vertex
$ gf^{abc}E^b_nC^cv_n $. The ghost propagator is $ {1\over{K^2}} $ and it is represented with the double line.
\vskip 0.5cm
\noindent
Fig.9b. Diagram with an open ghost line. The source $ u_i^a $ of the transverse gluon field has the vertex
$ gf^{abc}\delta_{ij} $.

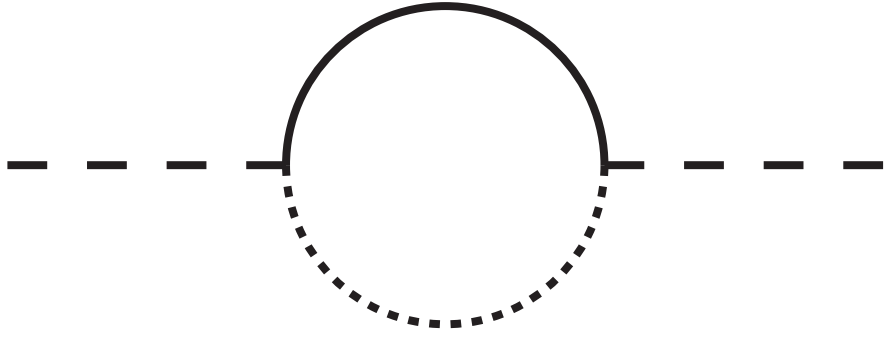
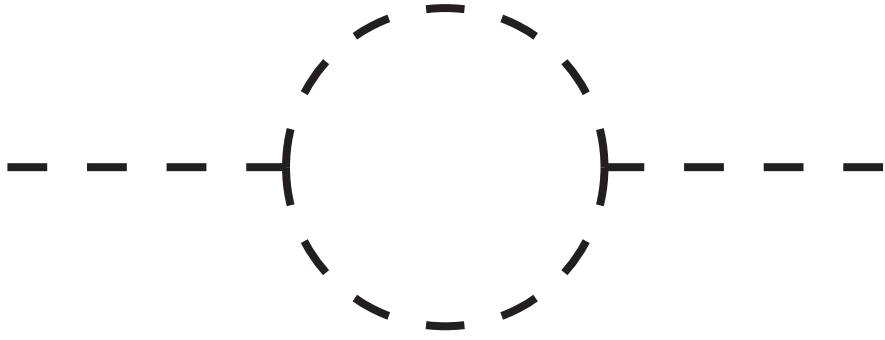
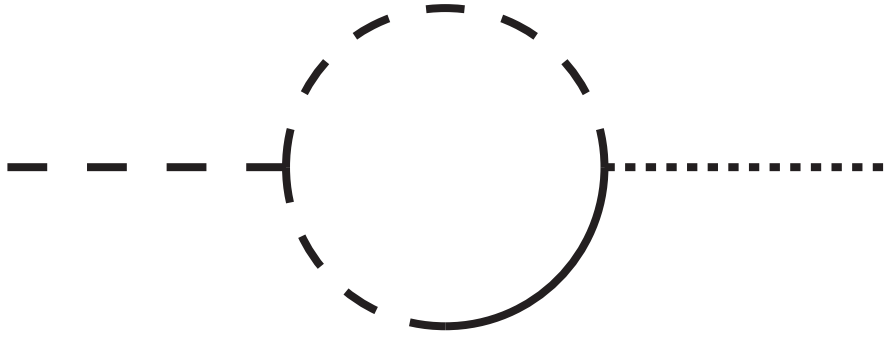
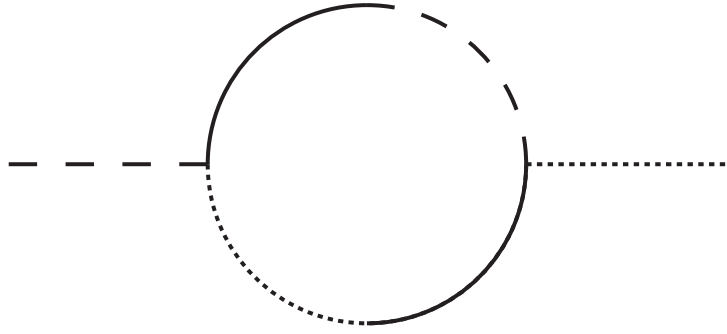
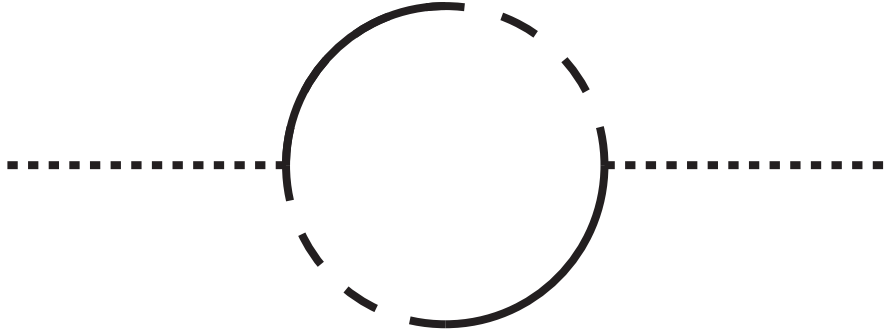
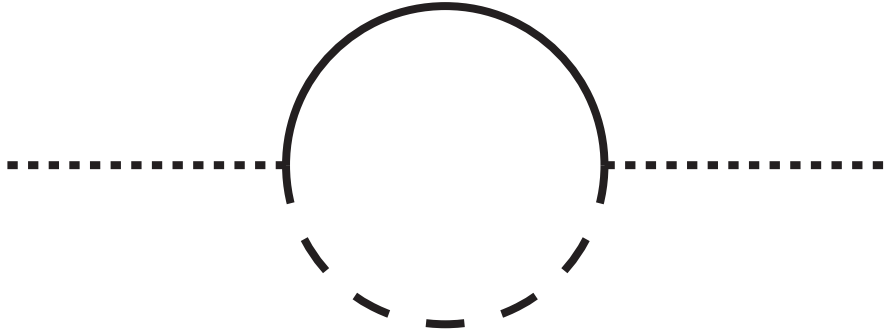
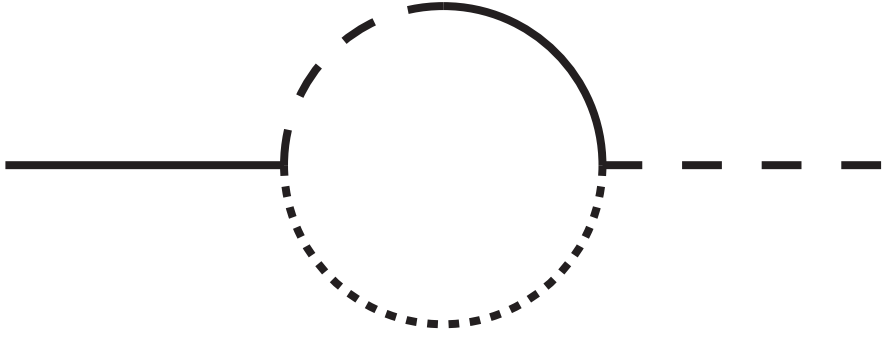
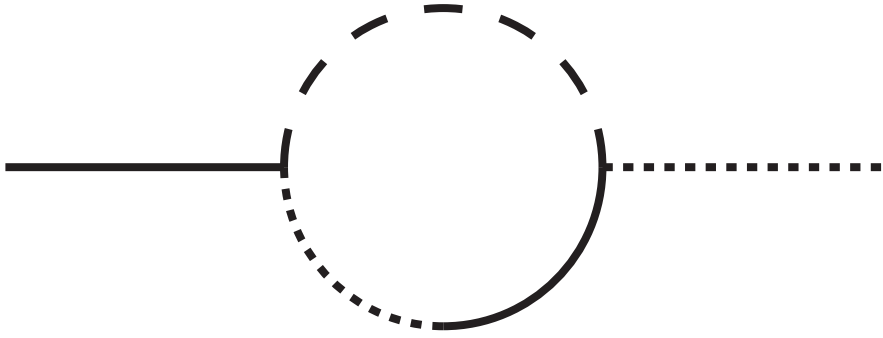
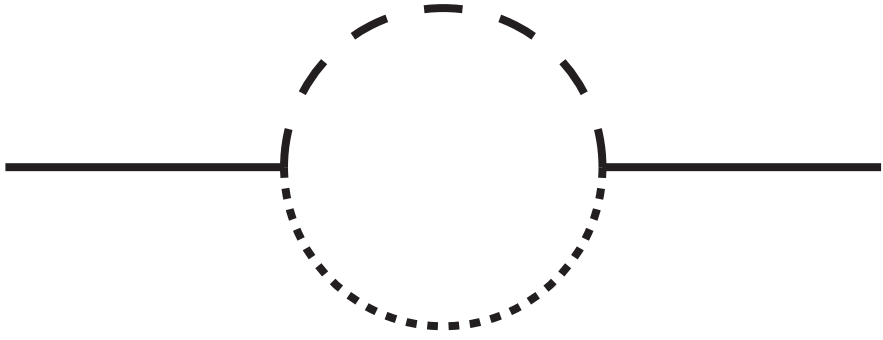
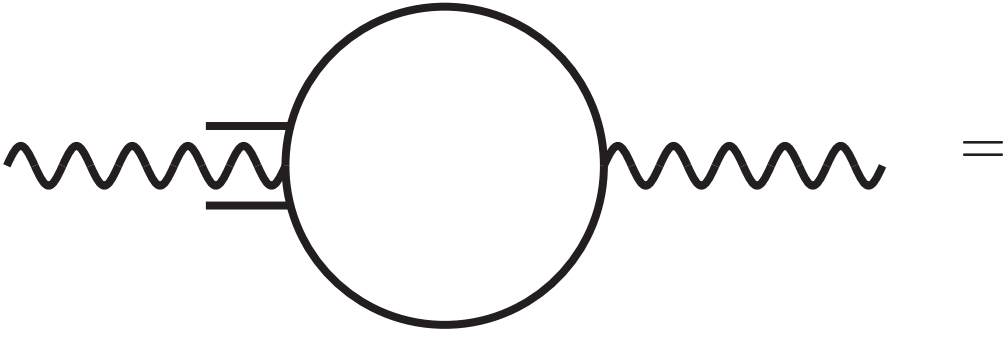
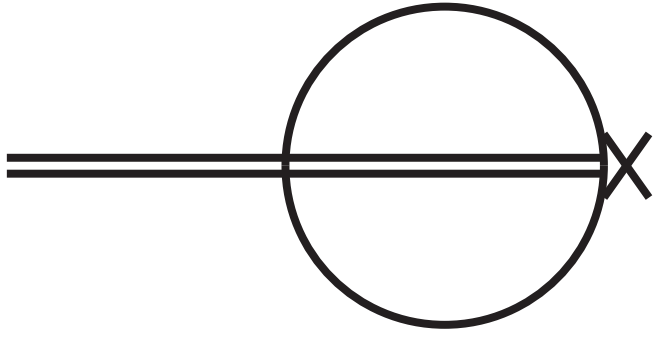
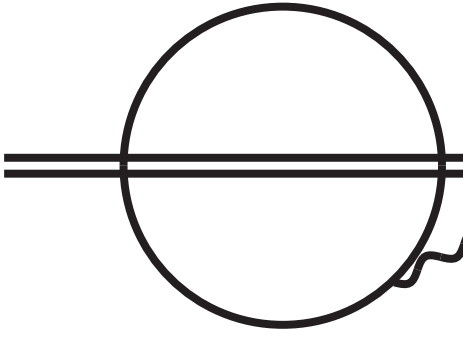
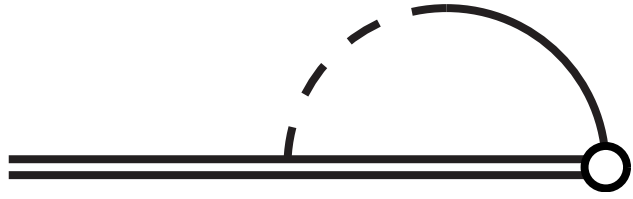
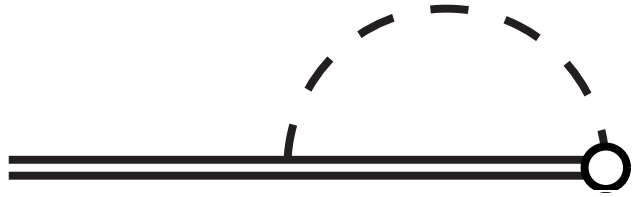
\bye